\documentclass[manuscript]{acmart}
\usepackage{multirow}
\usepackage{booktabs}
\usepackage{array}
\usepackage{graphicx}
\usepackage[table]{xcolor}

\AtBeginDocument{%
  }

\setcopyright{acmlicensed}
\copyrightyear{2018}
\acmYear{2018}
\acmDOI{XXXXXXX.XXXXXXX}
\acmConference[CHI '26]{Proceedings of the 2026 CHI Conference on Human Factors in Computing Systems}{April 13--17, 2026}{Barcelona, Spain}
\acmISBN{978-1-4503-XXXX-X/2018/06}

\begin{document}

\title{Why Some Seek AI, Others Seek Therapists: Mental Health in the Age of Generative AI}


\author{Junsang Park}
\email{junsangpark@ufl.edu}
\orcid{0000-0002-8274-4673}
\affiliation{%
 \institution{University of Florida}
 \city{Gainesville}
 \state{Florida}
 \country{USA}
}

\author{Sarah Brown}
\email{sarah.brown@ufl.edu}
\orcid{1234-5678-9012}
\affiliation{%
 \institution{University of Florida}
 \city{Gainesville}
 \state{Florida}
 \country{USA}
}

\author{Sharon Lynn Chu}
\email{slchu@ufl.edu}
\orcid{1234-5678-9012}
\affiliation{%
 \institution{University of Florida}
 \city{Gainesville}
 \state{Florida}
 \country{USA}
}


\begin{abstract}
As generative artificial intelligence (GAI) enters the mental health landscape, questions arise about how individuals weigh AI tools against human therapists. Drawing on the Health Belief Model (HBM), this study examined belief-based predictors of intention to use GAI and therapists across two populations: a university sample (N = 1,155) and a nationally representative adult sample (N = 651). Using repeated-measures ANOVA and LASSO regression, we found that therapists were consistently valued for emotional, relational, and personalization benefits, while GAI was favored for accessibility and affordability. Yet structural advantages alone did not predict adoption; emotional benefit and personalization emerged as decisive factors. Adoption patterns diverged across groups: students treated GAI as a complement, whereas national adults approached it as a substitute. Concerns about privacy and reliability constrained GAI use in both groups. These findings extend HBM to multi-modality contexts and highlight design implications for trustworthy, emotionally resonant digital mental health tools.
\end{abstract}


\begin{teaserfigure}
  \includegraphics[width=\linewidth]{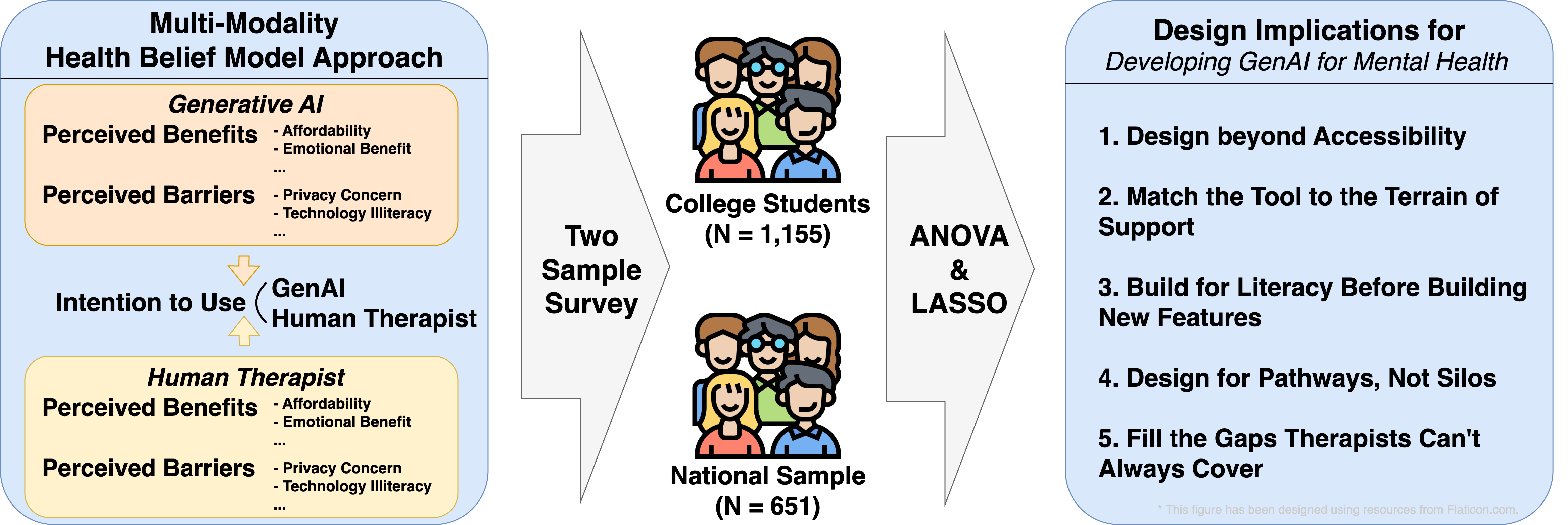}
  \caption[Overview of study design and implications]{Overview of the study design using a Health Belief Model approach across two samples, and resulting design implications for developing Generative AI (GenAI) for mental health.}
  \Description{Diagram summarizing study design and design implications for GenAI in mental health. Three horizontal panels.
  Left panel: Multi-Modality Health Belief Model Approach with two stacked boxes for Generative AI and Human Therapist, each listing perceived benefits (e.g., affordability, emotional benefit) and perceived barriers (e.g., privacy concern, technology illiteracy).
  Both feed into Intention to Use GenAI or Human Therapist.
  Middle panel: an arrow to Two Sample Survey with two participant icons labeled College Students (N=1,155) and National Sample (N=651).
  A second arrow points to analyses labeled ANOVA \& LASSO.
  Right panel: Design Implications for Developing GenAI for Mental Health listing:
  (1) Design beyond Accessibility;
  (2) Match the Tool to the Terrain of Support;
  (3) Build for Literacy Before Building New Features;
  (4) Design for Pathways, Not Silos;
  (5) Fill the Gaps Therapists Can’t Always Cover.
  Key takeaway: perceived benefits and barriers shape intentions to use GenAI versus human therapists across two samples, yielding design recommendations for future GenAI tools.}
  \label{fig:teaser}
\end{teaserfigure}

\received{20 February 2007}
\received[revised]{12 March 2009}
\received[accepted]{5 June 2009}

\maketitle

\section{Introduction}

Across the globe, digital technologies are becoming central to how individuals navigate mental health support.
While rates of anxiety, depression, and stress continue to rise, particularly among younger populations and those facing socioeconomic pressures, traditional care systems remain constrained by cost, stigma, and limited availability \cite{bor2014child, lipson2019increased, aguirre2020barriers}.
This gap has fueled interest in interactive tools—especially those powered by generative artificial intelligence (GAI)—that promise scalable, on-demand, and low-cost support.

Conversational agents such as ChatGPT have introduced new ways for people to access guidance, validation, and psychoeducational resources \cite{lattie2022overview, guo2024large, hua2024large}.
At the same time, many continue to rely on human therapists for relational depth, diagnostic nuance, and long-term growth.
Rather than replacing one another, these modalities are increasingly encountered side by side, creating a socio-technical ecosystem where users must navigate decisions between AI-driven and human-provided care.
This moment thus raises critical questions:
How do users perceive and evaluate the respective roles of GAI tools and human therapists?
Are these modalities viewed as complementary supports that can coexist, or as interchangeable options competing for preference?
And, from a design perspective, which aspects of GAI are perceived as most influential in shaping users’ intention to engage with such tools?

While prior work has examined barriers to therapy or the adoption of mental health apps, most studies treat each modality in isolation \cite{jardine2024between, naslund-2017, aguirre2020barriers}.
This fragmented approach limits our understanding of the comparative processes that drive real-world help-seeking decisions.
In practice, users often evaluate their options across modalities, weighing the strengths of GAI against the perceived limitations of therapy, or vice versa.
However, empirical research that systematically compares these decisions is lacking.
Our work draws on the Health Belief Model (HBM) as a guiding framework to study users' decision-making.
The HBM has been widely used to explain health-related behaviors by identifying how perceived benefits, perceived barriers, and contextual factors shape individuals' engagement with care \cite{janz-1984}.
By applying this model to both GAI and human therapy, we assess how users’ beliefs about each modality contribute to their intention to use it.
Rather than assuming individuals view mental health support as a single choice, we investigate whether GAI and human therapists are perceived as mutually exclusive options, where increased reliance on one leads to decreased use of the other, or as complementary supports, each offering unique benefits in different situations.
This perspective reflects how choices are often shaped not just by attraction to one option or dissatisfaction with another, but by how the options function together in a broader mental health ecosystem. 

Therefore, our study posed the following research questions:

\begin{itemize}
    \item \textbf{RQ1:} How do individuals' perceptions of GAI and human therapists differ across psychological dimensions (e.g., emotional support, personalization, trust)?
    \item \textbf{RQ2:} Do these perceptions vary across populations, and what population-specific patterns emerge when comparing a university sample to a national adult sample?
    \item \textbf{RQ3:} Do intentions to use GAI and therapists for mental health support influence one another? For example, if a person intends to use GAI, does that decrease their intention to seek support from a therapist?
    \item \textbf{RQ4:} What belief-based factors most strongly predict intention to use GAI versus human therapists?
\end{itemize}

By examining these questions across two distinct populations, a university student sample and a nationally representative adult sample, we are also able to explore how different populations have distinctive modality preference.
This study provides a side-by-side analysis of GAI and human therapists as socio-technical supports, advancing theoretical understanding of user decision-making and offering actionable design implications for the future development of mental health technologies. 
To summarize, our contributions include:
\begin{itemize}
    \item The extension of the Health Belief Model into a multimodality framework that captures how beliefs about GAI and human therapists interact in shaping help-seeking decisions,
    \item The identification of belief-based factors that most strongly predict intention to use each modality, revealed through LASSO regression across a university student sample and a national adult sample, and
    \item A set of design implications for developing GAI-enabled mental health tools that address individuals’ needs for GAI in relation to the mental helath care within the broader socio-technical ecosystem.
\end{itemize}

\section{Background and Related Work}

\subsection{Mental Health and Technology Adoption}

A growing body of HCI research has explored how digital technologies can help bridge gaps in mental health support.
Traditional barriers to accessing human therapists, including high costs, provider shortages, long wait times, and stigma, have driven many individuals toward more scalable, on-demand alternatives ~\cite{aguirre2020barriers, andrade2014barriers}.
In this context, mental health chatbots and generative AI (GAI) tools have gained traction as promising supplemental resources.
These tools simulate therapeutic conversations, deliver psychoeducational content, and offer behavioral exercises with 24/7 accessibility and reduced perceived judgment\cite{hua2024large, guo2024large, lattie2022overview}.
Notable examples include Woebot, applications that use scripted or AI-driven dialogue to guide users through mood tracking, cognitive reframing, and mindfulness practices ~\cite{fitzpatrick-2017}.

More recently, large language models (LLMs) such as ChatGPT have introduced open-ended conversational capabilities that users have begun informally repurposing for emotional support ~\cite{wang2025understanding, alanezi-2024}.
Despite limited clinical validation, recent surveys indicate that a significant proportion of people are experimenting with AI-based tools for mental health conversations ~\cite{zao-sanders-2025}.
These developments raise critical questions about user preferences, expectations, and beliefs when choosing between AI-mediated support and traditional human therapy, particularly as both modalities become more accessible and socially normalized.

\subsection{The Health Belief Model in Digital Mental Health}

The Health Belief Model (HBM)~\cite{janz-1984} is a widely used framework for explaining why individuals initiate or avoid health-related actions.
It posits that behavior is a function of core belief components: perceived \emph{susceptibility} (likelihood of experiencing a condition) and \emph{severity} (anticipated consequences), which jointly shape perceived threat; perceived \emph{benefits} (expected gains from the action) weighed against perceived \emph{barriers} (anticipated costs or obstacles); \emph{self-efficacy} (confidence in one’s ability to perform the action); and \emph{cues to action} (internal or external triggers that prompt behavior).
Although originally developed for preventive behaviors (e.g., vaccination, screening), the HBM has been adapted to help-seeking contexts where the target action is engaging with a service, adhering to treatment, or adopting a technology-supported health tool \cite{carpenter2010meta}.

Applied to mental health help-seeking from human clinicians, HBM constructs map naturally onto common determinants of therapy use \cite{henshaw2009conceptualizing}.
Perceived susceptibility and severity reflect a person’s recognition of symptoms and their impact; perceived benefits capture expectations about symptom relief, emotion regulation, or relational gains through therapeutic alliance; perceived barriers encompass direct and indirect costs (fees, time, travel), access constraints (waitlists, provider shortages), stigma, cultural mismatch, and confidentiality concerns; self-efficacy represents confidence in finding, initiating, and sustaining treatment; and cues to action include referral, screening outcomes, or acute stressors.
In digital mental health, the HBM has been used to explain uptake and continued engagement with mobile apps, telehealth, and web-based interventions~\cite{zhao-2022,naslund-2017,zahed-2023}.
Here, perceived benefits often include immediacy, convenience, personalization, and on-demand psychoeducation, while perceived barriers center on privacy/security risks, uncertainty about reliability or clinical effectiveness, low mental health literacy, and technology literacy or access constraints.
Across these settings, HBM-guided studies consistently show that higher perceived benefits and self-efficacy, together with manageable barriers, predict stronger intentions and greater use, whereas heightened perceived barriers and low self-efficacy suppress adoption and adherence~\cite{zhao-2022,naslund-2017,zahed-2023}.

A limitation of prior HBM applications in this domain is that they typically evaluate a \emph{single} support modality in isolation (e.g., therapists \emph{or} a particular digital tool), implicitly assuming that beliefs about one option do not affect intentions toward the other.
In practice, individuals often consider multiple, partially substitutable options simultaneously (e.g., a human therapist and a generative-AI tool) \cite{young2024role}.
Under such conditions, beliefs about one modality may carry \emph{cross-modal} consequences: for example, perceiving strong benefits of GAI (e.g., 24/7 access, anonymity, affordability) could \emph{reduce} intention to seek a therapist, while perceived barriers to therapy (e.g., cost, wait times, stigma) could \emph{increase} intention to use GAI; conversely, doubts about GAI’s effectiveness or data practices could \emph{increase} preference for human care.

\begin{figure}[ht]
  \includegraphics[width=\linewidth]{HBM-Figures.png}
  \caption[Single- vs. multi-modality HBM approach]
  {Comparison of a single-modality versus multi-modality Health Belief Model (HBM) approach for explaining intentions to use Generative AI (GAI) or a human therapist.
  The single-modality panel models benefits and barriers separately for GAI and for therapists with direct effects on each intention.
  The multi-modality panel adds cross-modality influences, where benefits and barriers for both GAI and therapists jointly increase or decrease intention to use GAI.}
  \Description{Diagram with two panels.
  Left: titled Single-Modality HBM Approach.
  Top section shows Perceived Benefits of GAI and Perceived Barriers of GAI with arrows pointing to an Intention to Use GAI icon.
  Blue arrows labeled Increase intention originate from benefits and red arrows labeled Decrease intention originate from barriers.
  A horizontal divider separates the bottom section, which mirrors the structure for human therapists: Perceived Benefits of Therapist and Perceived Barriers of Therapist point to Intention to Use Therapist with blue increase and red decrease arrows.
  Right: titled Multi-Modality HBM Approach.
  Four stacked constructs appear on the left: Perceived Benefits of GAI, Perceived Barriers of GAI, Perceived Benefits of Therapist, and Perceived Barriers of Therapist.
  From these, multiple slanted arrows point to a single outcome on the right labeled Intention to Use GAI.
  Blue arrows from benefits indicate increased intention; red arrows from barriers indicate decreased intention.
  Key takeaway: the multi-modality version models cross-source benefits and barriers together, capturing how perceptions of both GAI and therapists jointly shape intention to use GAI, beyond modeling each modality in isolation.}
  \label{fig:hbm}
\end{figure}

To address this gap, our study operationalizes parallel, modality-specific HBM constructs (multi-modulality HBM) for both GAI and human therapists and analyzes them \emph{jointly} (see Figure~\ref{fig:hbm}).
This design enables estimation of both \emph{within-modality} effects (e.g., how GAI benefits predict GAI intention) and \emph{cross-modality} effects (e.g., how therapist barriers predict GAI intention, and vice versa), offering a more complete account of how perceived benefits, barriers, and efficacy shape real-world mental health support choices.

\subsection{Perceived Benefits and Barriers of Therapists and Generative AI}
\label{sec:benefit_barrier}
\noindent\textbf{Emotional and Relational Support.}
One of the most commonly cited motivations for seeking mental health services is the need for emotional understanding and interpersonal connection~\cite{fluckiger2018alliance, fluckiger2020assessing}.
Human therapists are uniquely positioned to deliver empathic presence, attuned feedback, and relationship-based support.
These elements not only facilitate trust and vulnerability but also serve as active ingredients in therapeutic change.
The therapeutic alliance has been repeatedly shown to predict outcomes across modalities, especially for complex or identity-laden concerns~\cite{fluckiger2018alliance}.
In contrast, GAI tools can simulate empathic responses and offer immediate emotional validation, but their lack of lived experience and authentic presence limits the perceived depth of this connection~\cite{ovsyannikova2025third, montemayor2022principle}.
For some, GAI’s nonjudgmental nature may be a relief; for others, it may feel emotionally insufficient or uncanny~\cite{volpato2025trusting, seitz2024artificial}.

\noindent\textbf{Practical and Psychoeducational Needs.}
Beyond emotional support, individuals seek mental health tools for structured strategies, skill-building, and psychoeducation, especially when managing stress, anxiety, or depressive symptoms~\cite{park2025integrating, donker2009psychoeducation}.
Human therapists can offer highly tailored interventions, drawing from diverse clinical frameworks and adapting them to each client’s context.
This tailoring is supported by dialogue, intuition, and training.
GAI tools, on the other hand, excel in scalable delivery of structured content, such as mindfulness scripts, coping techniques, and explanations of psychological principles~\cite{inkster2018empathy}.
For users seeking immediate guidance or knowledge—especially in moments of distress—GAI can fulfill a practical function~\cite{park2025integrating, dehbozorgi2025application}.
Such tools may offer immediate guidance during moments of distress, but they remain limited in personalizing their responses based on complex clinical histories, comorbidities, or nuanced client context.

\noindent\textbf{Accessibility, Cost, and Anonymity.}
Structural barriers remain among the strongest deterrents to engaging with traditional therapy.
High costs, limited insurance coverage, long waitlists, and geographic shortages of providers are pervasive, especially for marginalized or rural populations~\cite{nadarzynski2025chatbot, coombs2021barriers}.
GAI tools mitigate many of these concerns: they are available 24/7, typically free or low-cost, and easily accessed via smartphone or computer~\cite{xian2024debate, sezgin2024behavioral}.
Moreover, for those who fear social stigma or feel uncomfortable disclosing sensitive issues to another person, GAI tools offer relative anonymity and control over disclosure pacing~\cite{xian2024debate, sezgin2024behavioral}.
These practical affordances position GAI as a promising alternative or supplement to traditional services—particularly for those who are not ready or able to seek human care.

\noindent\textbf{Trust, Reliability, and Literacy Barriers.}
Substantial barriers limit engagement with both GAI and therapists.
For therapists, trust may be impeded by past negative experiences, mismatches in cultural background, or fear of judgment~\cite{vybiral2024negative, gopalkrishnan2018cultural, ahmed2021mental}.
For GAI, skepticism often centers on reliability and privacy: can the tool offer helpful or safe responses?
How is data stored or used?
These concerns are heightened in sensitive domains like trauma or suicidality, where precision and accountability are critical~\cite{xian2024debate, sezgin2024behavioral}.
Additionally, both types of support may be limited by the user’s mental health or technology illiteracy, users may not recognize when they need help, or know how to initiate and navigate the support available~\cite{xian2024debate}.

\noindent\textbf{Tailoring These Dynamics to Benefit and Barrier Constructs.}
Collectively, these motivations and hesitations map onto the key benefit and barrier constructs assessed in this study.
Individuals may value emotional validation, practical guidance, personalization, relational support, psychoeducation, accessibility, and affordability—whether provided by humans or machines.
Conversely, their help-seeking behavior may be constrained by privacy concerns, doubts about reliability, stigma, cultural norms around self-reliance or professional help, and gaps in either mental health or technology illiteracy.
Importantly, different support modalities address these needs and concerns in distinct ways.
Our study builds on this understanding by examining how people evaluate the trade-offs between GAI and therapists, and how these evaluations predict intention to use one, both, or neither.

\section{Method}
All research procedures were approved by the Institutional Review Board (IRB) of the first author's affiliated institution.
Data collection took place between April and May 2025.
Participants in the university sample received course credit, while participants recruited via Prolific were compensated in accordance with platform standards (\$10.81/hour; estimated survey duration: 18–20 minutes).

\subsection{Participants}
\subsubsection{University Sample}
We conducted an online survey using Qualtrics and recruited participants through departmental research pools in psychology and computer science at a university located in the southeastern United States.
A total of 784 students were recruited from the computer science department and 482 from the psychology department, yielding an initial sample of 1,266 students.
Participants received one course credit upon completing the survey.
After excluding 111 participants who did not complete the survey, the final analytic sample consisted of 1,155 students.

The mean age of participants was 20.57 years (\textit{SD} = 4.06).
In terms of gender identity, 50.0\% identified as female, 47.4\% as male, and 2.6\% as nonbinary or another gender.
Regarding racial identity, the majority identified as White (35.06\%), followed by Asian (24.68\%) and Multiracial (18.70\%).
Hispanic participants comprised 9.70\% of the sample, and 4.42\% identified as Black or African American.
Smaller proportions identified as Middle Eastern or North African (1.39\%), Other (0.78\%), Native American or Alaska Native (0.17\%), and Native Hawaiian or Pacific Islander (0.17\%).
An additional 0.43\% selected ``Multiracial'' as a single-race category, and 4.50\% did not report racial identity.

In terms of educational attainment, 55.03\% reported having completed some college or an associate degree, while 26.32\% had a high school diploma or GED.
A smaller portion had completed a bachelor’s degree (9.37\%), and fewer had completed only some high school (2.58\%).
Additionally, 1.72\% held a master’s degree, and 0.52\% had earned a doctoral or professional degree (e.g., JD, MD, Ph.D.).

Household income levels varied across the sample.
The largest group (40.46\%) reported an annual household income of \$100{,}000 or more.
Other income brackets included less than \$20{,}000 (17.77\%), \$20{,}000–39{,}999 (9.18\%), \$40{,}000–59{,}999 (8.83\%), \$60{,}000–79{,}999 (8.31\%), and \$80{,}000–99{,}999 (8.23\%).

Regarding generative AI (GAI) use, participants were asked how frequently they engage with tools such as ChatGPT.
Most reported at least some use, with 33.99\% indicating ``sometimes,'' 22.86\% ``about half the time,'' 18.93\% ``most of the time,'' and 13.58\% ``always.''
A smaller proportion (7.10\%) reported never using such tools.

Participants also indicated the roles they believe GAI tools play in mental health support (select-all-that-apply format).
The most frequently endorsed role was as an informational resource (79.03\%).
Other common roles included motivational coach (22.02\%), friend (20.50\%), and emotional support (19.83\%).
A smaller subset (9.72\%) selected therapist substitute, indicating a more limited view of AI as a replacement for human therapists.

\subsubsection{National Sample}
We collected a national sample through an online survey administered via Qualtrics and distributed using Prolific, an online crowdsourcing platform commonly used for academic research.
Prolific enables targeted recruitment based on demographic criteria and provides built-in quality control measures, including pre-screening and attention checks.
A total of 702 U.S.-based adults were recruited, of which 51 were excluded for failing attention checks or not completing the survey, resulting in a final analytic sample of 651 participants.

Participants ranged in age from 18 to 83 years (\textit{M} = 45.62, \textit{SD} = 16.00).
The majority identified as White (61.54\%), followed by Black or African American (12.92\%) and Multiracial (8.62\%).
Hispanic participants comprised 6.46\% of the sample, while 5.38\% identified as Asian.
Smaller proportions identified as Native American or Alaska Native (1.54\%), Multiracial as a single-race category (1.08\%), Other (0.62\%), Native Hawaiian or Pacific Islander (0.31\%), and Middle Eastern or North African (0.15\%).
A total of 1.38\% of participants did not report their racial identity.

The sample was highly educated overall.
The largest proportion of participants had completed a bachelor’s degree (39.77\%), followed by those with some college or an associate degree (28.95\%) and a master’s degree (20.30\%).
Smaller proportions had completed a high school diploma or GED (12.35\%), a doctoral or professional degree (7.11\%), or only some high school (0.34\%).

Employment status varied across the sample.
Just over half were employed full-time (51.38\%), and an additional 25.86\% were employed part-time.
Other participants were retired (11.91\%), unemployed and seeking work (10.41\%), or unemployed and not seeking work (5.37\%).
A small number (2.69\%) were students not currently employed.

Participants reported a broad range of household income levels.
The largest group (29.23\%) reported an annual household income of \$100{,}000 or more.
Other commonly reported income brackets included \$20{,}000–39{,}999 (19.16\%), \$60{,}000–79{,}999 (18.97\%), and \$40{,}000–59{,}999 (18.14\%).
Smaller proportions reported earning \$80{,}000–99{,}999 (11.75\%) or less than \$20{,}000 (10.75\%).

In terms of AI engagement, participants were asked how frequently they use generative AI (GAI) tools such as ChatGPT.
The majority reported at least occasional use, with 51.25\% selecting “sometimes,” 21.36\% “most of the time,” 12.68\% “about half the time,” and 12.18\% “always.”
A minority (10.19\%) reported never using such tools.

Participants also selected all applicable roles they believe GAI tools play in supporting mental health.
The most frequently endorsed role was as an \textit{informational resource} (77.32\%).
Other commonly selected roles included \textit{motivational coach} (34.49\%), \textit{emotional support} (32.91\%), and \textit{therapist substitute} (29.76\%).
Additionally, 25.98\% identified GAI tools as serving the role of a \textit{friend}, reflecting perceived relational or affective dimensions of interaction with these technologies.

\subsection{Measures}
To guide our measurement development, we reviewed prior literature exploring the benefits and barriers of GAI- and therapist-based mental health support (see Section~\ref{sec:benefit_barrier}).
Across these sources, recurring themes included emotional trust, accessibility, personalization, stigma, and cost.
We also incorporated additional constructs—such as psychoeducation and mental health literacy—drawn from broader research on digital mental health engagement and barriers to care.
All items were written in a parallel comparison format—that is, the same items were instantiated twice to assess perceived benefits and barriers for each modality (once for GAI and once for therapists).
Where validated instruments existed, we adapted wording from established scales to preserve construct validity while enabling parallel, modality-specific comparisons (e.g., Barriers to Access to Care Evaluation (BACE) scale ~\cite{clement2012development} ;  eHealth Literacy ~\cite{norman2006ehealth}; psychometric properties of the credibility and expectancy ~\cite{devilly2000psychometric}.
However, most available measures are modality-bound (e.g., therapist-only or technology-only) and are not designed for side-by-side assessment across AI and human care.
We therefore created parallel item pairs that (1) retained the core meaning of validated measures and (2) could be administered identically for both GAI and therapists, supporting cross-modal analysis.
Developed items were then reviewed by HCI and counseling psychology researchers with 3+ years of domain experience to ensure clarity and coverage.

\textbf{Perceived Benefits.}
Participants were asked to indicate their level of agreement with a series of statements describing potential benefits of using either a human therapist or a generative artificial intelligence (GAI) tool for mental health support.
Perceived benefits were measured across 7 constructs: emotional benefit (e.g., feeling emotionally supported and validated), practical benefit (e.g., learning effective coping skills and strategies), relational benefit (e.g., improving one’s ability to manage relationships or social situations), educational benefit (e.g., receiving information in a clear and understandable format), personalization (e.g., tailored advice or interventions that meet individual needs), accessibility (e.g., immediate availability when needed), and financial affordability (e.g., being a more affordable option for mental health support).
Each construct was assessed using three items. Reliability analyses indicated acceptable internal consistency, with Cronbach’s alpha values exceeding 0.70 for all constructs.
For analysis, mean scores were computed for each construct.

\textbf{Perceived Barriers.}
Participants were asked to indicate their level of agreement with a series of statements describing potential barriers to using either a human therapist or a generative artificial intelligence (GAI) tool for mental health support.
Perceived barriers were measured across 5 constructs (6 for GAI): concern of privacy (e.g., discomfort with sharing sensitive or personal information), concern of reliability (e.g., doubts regarding the reliability or helpfulness of the support provided), concern of stigma (e.g., feeling embarrassed or ashamed about seeking mental health support), help seeking norm (e.g., preference for informal support from family or friends), mental health illiteracy (e.g., lack of awareness or understanding of available resources), and technology illiteracy (only for GAI; e.g., fear of technical issues or distrust in AI-generated responses).
Each construct was assessed using three items. Reliability analyses indicated acceptable internal consistency for all constructs, with Cronbach’s alpha values exceeding 0.70.
For analysis, mean scores were computed for each construct.

\textbf{Intention to Use.}
Participants’ intention to use each type of mental health support was assessed using two single-item measures: (1) “I intend to use GAI to support my mental health regularly,” and (2) “I intend to use a human professional to support my mental health regularly.”
Each item was rated on a continuous scale ranging from 0 to 100.
For analysis, mean scores were computed for each construct.

\subsection{Data Analysis}
To examine how individuals evaluate generative AI (GAI) and human therapists for mental health support, we employed a three-part analytic strategy integrating both inferential and predictive approaches.

First, we conducted repeated-measures ANOVA (RM-ANOVA) to profile how participants perceive the benefits and barriers of each support modality.
This analysis allowed us to determine whether the same type of benefit (e.g., emotional support, personalization) or barrier (e.g., privacy concern, cost) was rated differently when attributed to GAI versus therapists.
Post hoc comparisons were conducted to examine pairwise differences and identify which specific benefit or barrier domains diverged across modalities.
This step established a foundational understanding of the perceived strengths and weaknesses of each modality.

Second, we assessed the overall relationship between the two support types by regressing intention to use GAI on intention to use human therapists, controlling for key covariates including age, gender, previous GAI use, and previous therapy experience.
This analysis enabled us to explore whether GAI and therapist support are perceived as substitutes or complements.
A negative association suggests a substitution pattern, where greater intention to use one modality predicts lower intention to use the other.
Conversely, a non-significant or positive association implies a complementary relationship.

Third, to identify the specific belief-based factors most predictive of intention to use each modality, we employed Least Absolute Shrinkage and Selection Operator (LASSO) regression.
LASSO is a regularized regression technique that constrains the sum of the absolute values of the model coefficients, simultaneously performing variable selection and model regularization ~\cite{tibshirani1996regression}.
This approach is particularly well-suited to models with many potentially correlated predictors, such as the multiple perceived benefits and barriers measured in this study~\cite{zhou2024comparison, mcneish2015using}.

Separate LASSO regressions were conducted for two dependent variables: intention to use GAI and intention to use a human therapist.
Analyses were also conducted independently for the university and national samples to identify population-specific patterns.
Each model included all belief-based constructs (benefits and barriers) associated with both modalities.
This setup allowed us to capture both within-modality predictors (e.g., how beliefs about GAI predict GAI use) and cross-modality influences (e.g., how barriers to therapist use predict intention to use GAI).
LASSO models were implemented using the \texttt{glmnet} package in R. 

\section{Results}
In this section, we present the findings in the order of the four research questions.

\subsection{RQ1: How do individuals' perceptions of GAI and human therapists differ across benefits and barriers?}
Initial descriptive analyses examined trends in perceived benefits and barriers for Generative AI (GAI) versus human therapists across two distinct samples (national and university).

\begin{table}[ht]
  \caption[Means and SDs of perceived benefits and barriers]
  {Means and standard deviations of perceived benefits and barriers for mental health support, separately for Generative AI (GAI) and human therapists, in the national and university samples.
  Higher scores indicate stronger endorsement.}
  \Description{Table of mean scores and standard deviations for perceived benefits and barriers of Generative AI and human therapists, separated by national and university samples.
  Benefits include emotional, practical, relational, educational, personalization, accessibility, and affordability.
  Barriers include privacy, reliability, stigma, help-seeking norm, mental health illiteracy, and technology illiteracy.
  Intention to use is also reported. Key takeaway: GAI is rated higher on accessibility and affordability, while therapists score higher on emotional, relational, and practical benefits.
  Privacy and reliability are concerns for both, while stigma and literacy barriers vary by modality.}
  \label{tab:mental_health_support}
  \begin{tabular}{llcccc}
    \toprule
    \multicolumn{2}{c}{} & \multicolumn{2}{c}{National Sample} & \multicolumn{2}{c}{University Sample} \\
    \cmidrule(lr){3-4} \cmidrule(lr){5-6}
    Type & Construct & GAI & Therapist & GAI & Therapist \\
    \midrule
    \multirow{7}{*}{Benefit}
    & \cellcolor[HTML]{F2F2F2}Emotional Benefit & \cellcolor[HTML]{F2F2F2}53.15 (27.87) & \cellcolor[HTML]{F2F2F2}74.68 (19.64) & \cellcolor[HTML]{F2F2F2}45.47 (25.34) & \cellcolor[HTML]{F2F2F2}73.39 (20.71) \\
    & Practical Benefit & 60.65 (26.15) & 73.16 (20.58) & 53.90 (24.71) & 72.89 (21.07) \\
    & \cellcolor[HTML]{F2F2F2}Relational Benefit & \cellcolor[HTML]{F2F2F2}51.15 (27.52) & \cellcolor[HTML]{F2F2F2}72.39 (21.05) & \cellcolor[HTML]{F2F2F2}39.46 (25.55) & \cellcolor[HTML]{F2F2F2}72.13 (22.01) \\
    & Educational Benefit & 66.92 (24.67) & 73.64 (20.12) & 60.66 (25.53) & 72.22 (21.24) \\
    & \cellcolor[HTML]{F2F2F2}Personalization & \cellcolor[HTML]{F2F2F2}60.54 (27.28) & \cellcolor[HTML]{F2F2F2}73.39 (20.74) & \cellcolor[HTML]{F2F2F2}52.74 (26.30) & \cellcolor[HTML]{F2F2F2}73.19 (21.30) \\
    & Accessibility & 82.72 (21.14) & 52.57 (25.59) & 82.83 (23.29) & 49.76 (24.14) \\
    & \cellcolor[HTML]{F2F2F2}Affordability & \cellcolor[HTML]{F2F2F2}76.25 (24.94) & \cellcolor[HTML]{F2F2F2}50.81 (29.79) & \cellcolor[HTML]{F2F2F2}80.70 (25.00) & \cellcolor[HTML]{F2F2F2}44.64 (27.84) \\
    \midrule
    \multirow{6}{*}{Barrier}
    & Concern of Privacy & 65.67 (31.12) & 57.03 (29.13) & 70.59 (29.06) & 51.82 (28.02) \\
    & \cellcolor[HTML]{F2F2F2}Concern of Reliability & \cellcolor[HTML]{F2F2F2}62.65 (28.46) & \cellcolor[HTML]{F2F2F2}56.66 (26.09) & \cellcolor[HTML]{F2F2F2}69.44 (26.60) & \cellcolor[HTML]{F2F2F2}48.96 (26.82) \\
    & Concern of Stigma & 29.25 (27.74) & 49.36 (30.16) & 38.37 (29.42) & 46.77 (27.90) \\
    & \cellcolor[HTML]{F2F2F2}Help Seeking Norm & \cellcolor[HTML]{F2F2F2}32.60 (26.53) & \cellcolor[HTML]{F2F2F2}44.75 (28.44) & \cellcolor[HTML]{F2F2F2}40.64 (26.61) & \cellcolor[HTML]{F2F2F2}45.39 (26.64) \\
    & Mental Health Illiteracy & 40.80 (27.81) & 48.43 (27.97) & 41.63 (28.95) & 46.77 (26.96) \\
    & \cellcolor[HTML]{F2F2F2}Technology Illiteracy & \cellcolor[HTML]{F2F2F2}53.42 (28.37) & \cellcolor[HTML]{F2F2F2}- (-) & \cellcolor[HTML]{F2F2F2}56.37 (28.46) & \cellcolor[HTML]{F2F2F2}- (-) \\
    \midrule
    \multirow{1}{*}{Intention}
    & Intention to Use & 36.61 (31.42) & 58.54 (28.63) & 26.04 (27.44) & 55.79 (28.85) \\
    \bottomrule
  \end{tabular}
\end{table}

Table~\ref{tab:mental_health_support} presents the mean ratings and standard deviations for each benefit and barrier type, separated by source of support within each sample.
To examine overall perceptions, we conducted repeated-measures ANOVAs with benefit (or barrier) type as a within-subjects factor, collapsing across the university and national samples.

\textbf{Therapist perceptions.} 
The main effect of benefit type was significant, $F(6, 10296) = 935.93$, $p < .001$, indicating substantial differences across benefit categories. 
Participants rated therapists significantly higher on emotional, practical, relational, educational, and personalized benefits compared to accessibility and affordability. 
For example, emotional benefit was rated substantially higher than accessibility ($M_{\text{diff}} = 23.02$, $SE = 0.63$, $t(1716) = 36.73$, $p < .001$) and affordability ($M_{\text{diff}} = 26.57$, $SE = 0.76$, $t(1716) = 35.03$, $p < .001$). 
Personalization was similarly favored over affordability ($M_{\text{diff}} = 25.67$, $SE = 0.74$, $t(1716) = 34.57$, $p < .001$) and accessibility ($M_{\text{diff}} = 22.12$, $SE = 0.61$, $t(1716) = 36.09$, $p < .001$). 
Minor differences among the higher-rated benefits (e.g., personalization vs. practical) were statistically nonsignificant. 
Overall, therapists were valued most for emotional, relational, practical, and educational support, with accessibility and affordability seen as weaker dimensions.  

For barriers, the main effect of barrier type was also significant, $F(4, 6648) = 75.95$, $p < .001$. 
Privacy concerns emerged as the most prominent barrier, rated significantly higher than stigma ($M_{\text{diff}} = 6.67$, $SE = 0.68$, $t(1662) = 9.84$, $p < .001$), mental health illiteracy ($M_{\text{diff}} = 7.13$, $SE = 0.75$, $t(1662) = 9.50$, $p < .001$), reliability ($M_{\text{diff}} = 1.82$, $SE = 0.63$, $t(1662) = 2.91$, $p < .05$), and help-seeking norm ($M_{\text{diff}} = 9.60$, $SE = 0.73$, $t(1662) = 13.17$, $p < .001$). 
Help-seeking norm was the least endorsed barrier overall.  

\textbf{GAI perceptions.} 
For GAI, the main effect of benefit type was significant, $F(6, 10032) = 1165.20$, $p < .001$. 
Participants most strongly endorsed accessibility and affordability. 
Accessibility was rated significantly higher than affordability ($M_{\text{diff}} = 4.40$, $SE = 0.45$, $t(1672) = 9.71$, $p < .001$), personalization ($M_{\text{diff}} = 26.13$, $SE = 0.65$, $t(1672) = 40.25$, $p < .001$), and psychoeducation ($M_{\text{diff}} = 18.97$, $SE = 0.57$, $t(1672) = 33.56$, $p < .001$). 
In contrast, emotional and relational support were rated substantially lower (e.g., accessibility vs. emotional support: $M_{\text{diff}} = 33.28$, $SE = 0.70$, $t(1672) = 47.52$, $p < .001$). 
GAI also received moderate endorsement for practical and educational benefits.  

For barriers, the main effect of barrier type was significant, $F(5, 7885) = 747.15$, $p < .001$. 
Privacy and reliability concerns were most strongly endorsed, significantly higher than stigma, help-seeking norm, and technology illiteracy. 
For example, privacy concerns exceeded stigma ($M_{\text{diff}} = 34.07$, $SE = 0.93$, $t(1577) = 36.61$, $p < .001$) and help-seeking norms ($M_{\text{diff}} = 31.49$, $SE = 0.92$, $t(1577) = 34.39$, $p < .001$).  

Taken together, RQ1 results suggest that therapists are primarily valued for interpersonal and psychological benefits, while GAI is valued for structural advantages. 
Across both modalities, privacy and reliability consistently emerged as leading barriers.  

\begin{figure}[ht]
  \centering
  \includegraphics[width=\linewidth]{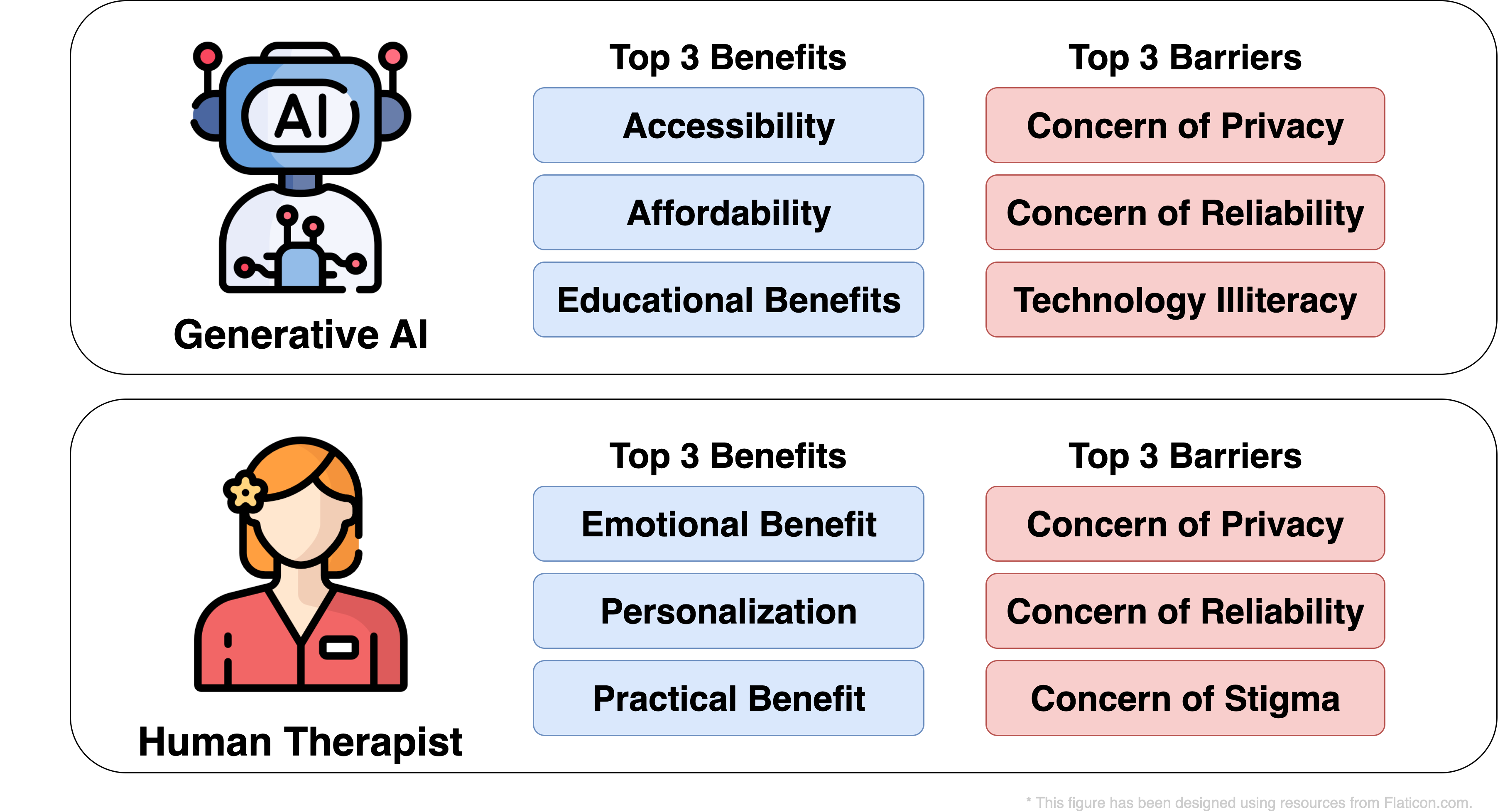}
  \caption[Profiles of perceived benefits and barriers]
  {Profiles of the top three perceived benefits and barriers for Generative AI (GAI) and human therapists, aggregated across the combined university and national samples ($N=1{,}806$).
  The figure shows that GAI is most valued for accessibility, affordability, and educational benefits, but is limited by privacy, reliability, and technology literacy concerns.
  Human therapists are valued for emotional benefit, personalization, and practical benefit, but face barriers of privacy, reliability, and stigma.}
  \Description{Two horizontal panels. Top panel: Generative AI profile.
  On the left is an icon of a robot labeled Generative AI.
  To the right, two columns list the top three benefits and barriers.
  Benefits (in blue boxes): Accessibility, Affordability, Educational Benefits.
  Barriers (in red boxes): Concern of Privacy, Concern of Reliability, Technology Illiteracy.
  Bottom panel: Human Therapist profile. On the left is an icon of a therapist labeled Human Therapist.
  Benefits (blue boxes): Emotional Benefit, Personalization, Practical Benefit.
  Barriers (red boxes): Concern of Privacy, Concern of Reliability, Concern of Stigma.
  Key takeaway: aggregated results across both samples highlight different benefit–barrier profiles for AI versus human therapists.}
  \label{fig:profiles}
\end{figure}

Figure~\ref{fig:profiles} summarizes the profiles of perceived benefits and barriers across the combined samples. 
These results reflect the mean scores of each benefit and barrier, collapsed across the university and national samples (N = 1806).

\subsection{RQ2: Do these perceptions vary across populations?}

We next examined group effects and group-by-benefit/barrier interactions to identify population-specific differences between the national and university samples.  

\textbf{Therapist perceptions.} 
The main effect of group was not significant for benefits, $F(1, 1716) = 2.63$, $p = .105$, suggesting that both samples valued therapists similarly across dimensions. 
However, group differences emerged for barriers: national participants reported significantly greater concerns about therapist reliability ($M_{\text{diff}} = 7.04$, $SE = 1.33$, $t(1662) = 5.28$, $p < .001$) and privacy ($M_{\text{diff}} = 4.50$, $SE = 1.42$, $t(1662) = 3.17$, $p < .01$) than university participants.  

\textbf{GAI perceptions.} 
For GAI, the main effect of group was significant, $F(1, 1672) = 22.92$, $p < .001$, as was the group-by-benefit interaction, $F(6, 10032) = 44.28$, $p < .001$. 
National participants rated GAI higher on relational ($M_{\text{diff}} = 11.85$, $SE = 1.32$, $t(1672) = 8.98$, $p < .001$), emotional ($M_{\text{diff}} = 7.04$, $SE = 1.32$, $t(1672) = 5.35$, $p < .001$), and educational benefits ($M_{\text{diff}} = 6.07$, $SE = 1.26$, $t(1672) = 4.82$, $p < .001$) than university participants.  

For barriers, the main effect of group was also significant, $F(1, 1577) = 23.65$, $p < .001$, as was the interaction, $F(5, 7885) = 9.62$, $p < .001$. 
University students consistently reported higher perceived barriers than national participants. 
This included greater concerns about stigma ($M_{\text{diff}} = 9.24$, $SE = 1.47$, $t(1577) = 6.28$, $p < .001$), help-seeking norms ($M_{\text{diff}} = 7.75$, $SE = 1.35$, $t(1577) = 5.73$, $p < .001$), and ineffectiveness ($M_{\text{diff}} = 6.31$, $SE = 1.40$, $t(1577) = 4.51$, $p < .001$).  

In sum, RQ2 results highlight population-specific patterns: national adults were more favorable toward GAI’s relational and emotional potential, while university students expressed stronger concerns about barriers, particularly privacy, reliability, and stigma. 
For therapists, both groups valued benefits similarly, but national adults reported greater sensitivity to privacy and reliability concerns.  

\begin{table*}[htbp]
\centering
\caption[Predictors of intention to use GAI and therapists]
{Perceived benefits and barriers predicting intention to use Generative AI (GAI) and therapist support for mental health across the university ($N=1{,}155$) and national ($N=651$) samples, based on LASSO regression coefficients.
Larger coefficients indicate stronger influence; blank cells reflect predictors not selected by the model.}
\Description{Table of LASSO regression coefficients showing which perceived benefits and barriers predict intention to use GAI or a therapist in university and national samples.
For GAI, relational and emotional benefits are strong positive predictors in the university sample, while emotional benefit and personalization dominate in the national sample.
Barriers such as reliability concerns and technology illiteracy reduce intention for GAI.
For therapists, emotional and affordability benefits are influential, especially in the national sample, while barriers like help‐seeking norms and privacy concerns lower intention.
R-squared values are 0.42 and 0.28 for university GAI and therapist models, and 0.52 and 0.29 for national models.
Key takeaway: benefits related to emotional connection and personalization drive intention, while reliability, literacy, and norms act as barriers.}
\label{tab:lasso_results}
\begin{tabular}{lllcccc}
\toprule
\multirow{3}{*}{\textbf{Type}} & \multirow{3}{*}{\textbf{Category}} & \multirow{3}{*}{\textbf{Predictor}} & 
\multicolumn{2}{c}{\textbf{University Sample}} & \multicolumn{2}{c}{\textbf{National Sample}} \\
\cmidrule(lr){4-5} \cmidrule(lr){6-7}
& & & \textbf{GAI} & \textbf{Therapist} & \textbf{GAI} & \textbf{Therapist} \\
\midrule

\multirow{7}{*}{GAI} & \multirow{7}{*}{Benefit}
& Emotional Benefit         & 0.19 & -0.04 & 0.40 & -0.03 \\
& & Practical Benefit              &      &        & 0.04 &       \\
& & Relational Benefit           & 0.21 &  -0.00 & 0.06 &       \\
& & Educational Benefit              &       &        &       &       \\
& & Personalization              & 0.17 &        & 0.16 & -0.01 \\
& & Accessibility                &       &  -0.03  &       & -0.03 \\
& & Affordability      & -0.05  & -0.01  &       & -0.06 \\
\specialrule{.1em}{.05em}{.05em}  

\multirow{6}{*}{GAI} & \multirow{6}{*}{Barrier}
& Concern of Privacy              &       & 0.07  & -0.03 & 0.06 \\
& & Concern of Reliability & -0.18 &       & -0.12 & 0.00 \\
& & Concern of Stigma                       & 0.02 &        & 0.07 & 0.03 \\
& & Help Seeking Norm            & 0.11 &        &       &       \\
& & Mental Health Illiteracy       &       &        &       & 0.08 \\
& & Technology Illiteracy          &      & 0.23 & -0.08 & 0.14 \\
\specialrule{.1em}{.05em}{.05em}  

\multirow{7}{*}{Therapist} & \multirow{7}{*}{Benefit}
& Emotional Benefit         & -0.09 & 0.15 & -0.03 & 0.33 \\
& & Practical Benefit              & -0.04  & 0.06 &       &      \\
& & Relational Benefit           &       & 0.08 &       & 0.04 \\
& & Educational Benefit              & -0.03 &       &       &       \\
& & Personalization              & -0.00 & 0.11 & -0.06 & 0.03  \\
& & Accessibility                &      &       & 0.02 & 0.01 \\
& & Affordability      & -0.00 & 0.22 &       & 0.24 \\
\specialrule{.1em}{.05em}{.05em}  

\multirow{5}{*}{Therapist} & \multirow{5}{*}{Barrier}
& Concern of Privacy              &       & -0.04 & 0.01 & -0.05  \\
& & Concern of Reliability & 0.11 & -0.04 &      & -0.01 \\
& & Concern of Stigma                       &       & -0.02  & 0.02 &       \\
& & Help Seeking Norm            & 0.01 &       & 0.09 & -0.13 \\
& & Mental Health Illiteracy       & 0.05 &       & 0.06 & 0.02 \\
\midrule
\multicolumn{3}{r}{\textbf{R\textsuperscript{2}}} & \textbf{0.42} & \textbf{0.28} & \textbf{0.52} & \textbf{0.29} \\
\bottomrule
\end{tabular}
\end{table*}

\subsection{RQ3: Do intentions to use GAI and therapists influence one another?}
We also examined how intention to use one modality influences intention to use the other.
To do so, we conducted multivariate regression analyses in both the national and university samples, controlling for age, gender, and prior experience with each modality.
In the national sample, a weak substitution effect was observed: intention to use GAI negatively predicted intention to use a therapist ($b = -0.10$, $SE = 0.04$, $p < .05$), and intention to use a therapist negatively predicted intention to use GAI ($b = -0.09$, $SE = 0.04$, $p < .05$).
This suggests that individuals in this sample tended to view the two modalities as alternatives, with greater interest in one associated with reduced interest in the other.
In contrast, the university sample showed no such pattern: intention to use GAI was not significantly related to intention to use a therapist ($b = 0.02$, $SE = 0.04$, $p = .44$), nor was intention to use a therapist significantly related to intention to use GAI ($b = 0.02$, $SE = 0.03$, $p = .44$).
This indicates a more complementary perception, where participants may be open to engaging with both GAI and human therapists simultaneously.

\subsection{RQ4: What perceived benefits/barriers most strongly predict intention to use GAI/human therapists?}

We then examined which psychological beliefs most strongly predict individuals’ intentions to use each modality by applying LASSO regression.
This method allowed us to identify the most influential predictors from a comprehensive set of perceived benefits and barriers, entered simultaneously for both GAI and therapist support.
By isolating the factors most predictive of intention for each modality, we gain insight into the specific drivers of use and how GAI tools can be designed to leverage the intersecting benefits and barriers of each modality to maximize adoption in the mental health landscape.
Standardized beta coefficients ($\beta$) are reported to enable interpretation and comparison across predictors.
Full LASSO coefficients are presented in Table~\ref{tab:lasso_results}, with corresponding visualizations provided in each results subsection.

\subsection*{University Sample}

\begin{figure}[ht]
  \centering
  \includegraphics[width=\linewidth]{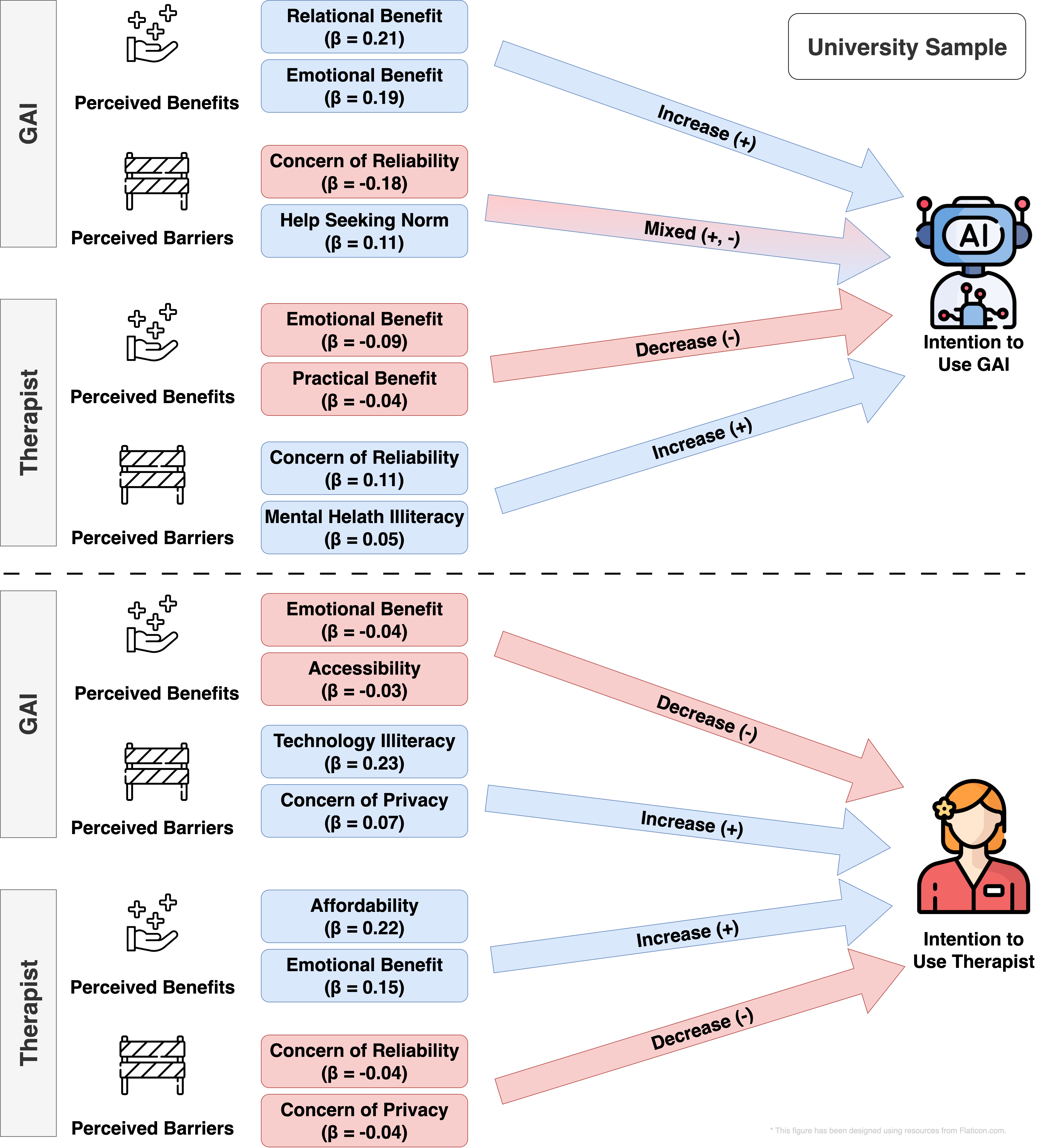}
  \caption[University sample LASSO results]
  {Summary of influential perceived benefits and barriers predicting intention to use Generative AI (GAI) or a human therapist in the university sample ($N=1{,}155$).
  For visibility, only the two largest coefficients in each category are shown.
  Full results are reported in Table~\ref{tab:lasso_results}.}
  \Description{Diagram divided into two halves, one for GAI (top) and one for human therapists (bottom), with arrows pointing toward the intention outcome.
  Top half (GAI): perceived benefits include Relational Benefit (beta=0.21) and Emotional Benefit (beta=0.19), shown as blue arrows labeled Increase.
  Perceived barriers include Concern of Reliability (beta=-0.18) shown as a red arrow labeled Decrease, and Help Seeking Norm (beta=0.11) shown as a mixed arrow labeled Mixed.
  Bottom half (Therapist): perceived benefits include Emotional Benefit (beta=-0.09) and Practical Benefit (beta=-0.04), both red arrows labeled Decrease.
  Perceived barriers include Concern of Reliability (beta=0.11) and Mental Health Illiteracy (beta=0.05), both blue arrows labeled Increase.
  Right side shows icons representing intention to use GAI (robot icon) and intention to use a therapist (person icon).
  Key takeaway: in the university sample, specific relational and emotional benefits drive GAI uptake, while therapist intention is more sensitive to reliability concerns and literacy barriers.}
  \label{fig:lasso_uni}
\end{figure}

Figure \ref{fig:lasso_uni} illustrates the significant perceived benefits and barriers for GAI and human therapist in the university sample. 

For GAI, three benefits of GAI emerged as significant facilitators.
Emotional benefit ($\beta = .19$), relational benefit ($\beta = .21$), and personalization ($\beta = .17$) were positively associated with students’ intention to use GAI for mental health support.
These findings suggest that students value emotionally responsive, personalized, and socially attuned interactions with GAI systems.
However, concern about reliability ($\beta = -0.18$) stood out as the strongest negative predictor, highlighting skepticism toward the intention to use GAI tools.
Notably, one of the most compelling insights lies in how beliefs about therapists influence GAI usage.
Students who were concerned about the reliability of human therapists were more likely to turn to GAI ($\beta = .11$), suggesting that perceived weaknesses in one modality may increase interest in the other.

Interestingly, a counterintuitive result emerged for the “help-seeking norm” barrier, which had a positive coefficient ($\beta = .11$), suggesting that students who perceive greater normative barriers to formally seeking help from GAI may still turn to GAI.
One possible interpretation is that students who are hesitant to engage with GAI may have higher barriers toward human therapists, so they may view GAI as a casual, low-stakes alternative. A similar dynamic may explain the weak but negative effect of affordability ($\beta = -0.05$): students who view GAI as highly affordable may actually have access to a broader range of support options including human therapist, thereby reducing their exclusive reliance on GAI. 

As for the intention to use therapists, students were more likely to engage when they perceived emotional benefit ($\beta = .15$), practical benefit ($\beta = .06$), relational benefit ($\beta = .08$), and personalization ($\beta = .11$) from therapy.
These findings reflect the multifaceted value students place on traditional therapeutic relationships.
Importantly, affordability was also a strong predictor ($\beta = .22$), underscoring the central role of financial access in the decision to pursue therapy, a factor that did not significantly predict GAI use in this sample.
Barriers to using therapy, however, were more limited in their influence. Additional barriers such as privacy concerns ($\beta = 0.07$) and technology illiteracy ($\beta = 0.23$) for using GAI also contributed to students’ intention to engage with therapist.
Conversely, therapist reliability concerns had a weak negative effect ($\beta = -0.04$), while privacy concerns ($\beta = -0.04$) showed similarly minimal influence. 

In sum, university students appear to adopt a pragmatic stance toward mental health support.
They turn to GAI when it offers emotional and relational engagement and when they question the reliability of traditional services.
However, concerns about GAI’s own reliability remain critical obstacles.
On the other hand, students prefer therapists when the relationship offers strong emotional and practical benefits and when services are financially accessible.

\subsection*{National Sample}
\begin{figure}[ht]
  \centering
  \includegraphics[width=\linewidth]{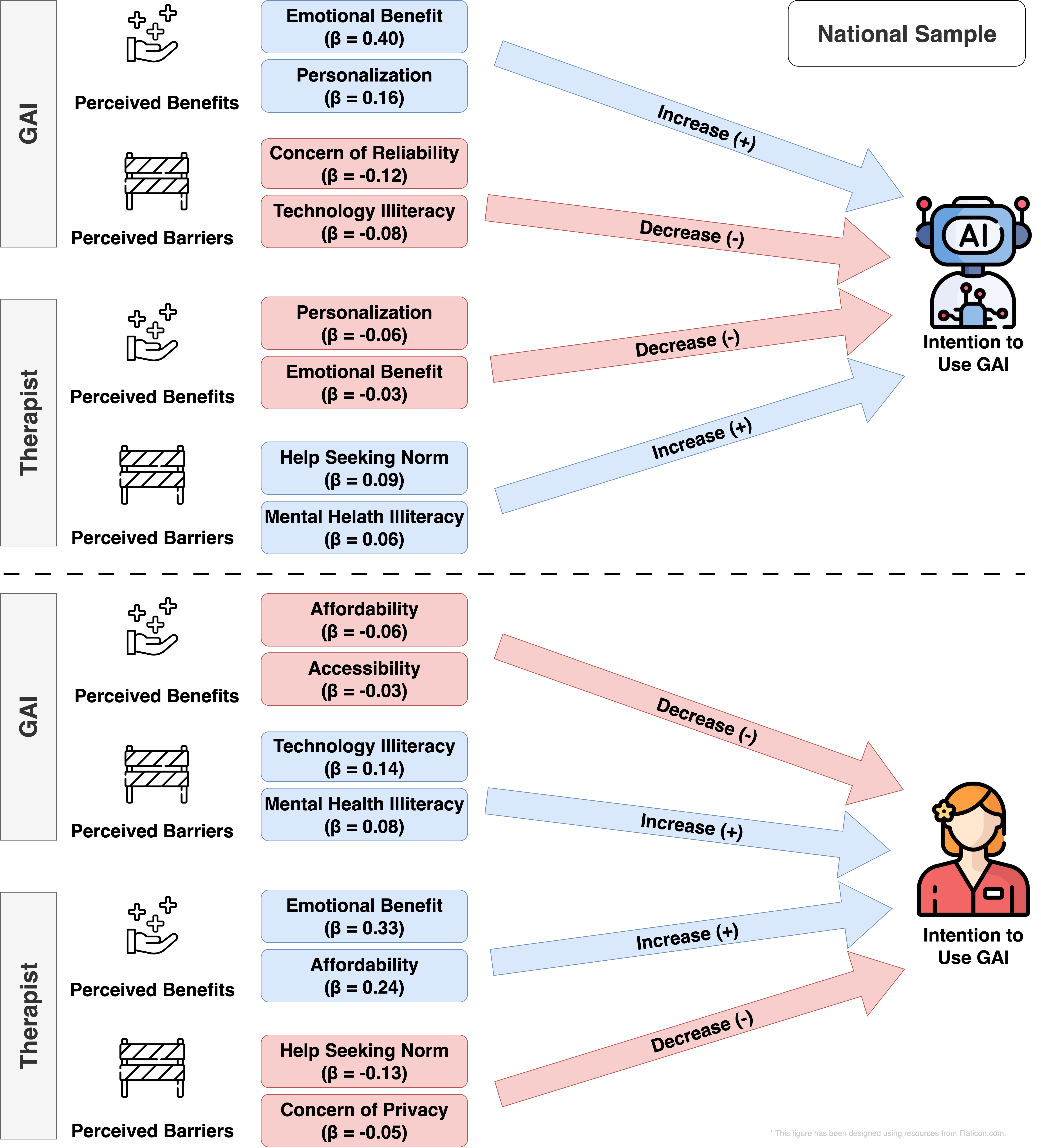}
  \caption[National sample LASSO results]
  {Summary of influential perceived benefits and barriers predicting intention to use Generative AI (GAI) or a human therapist in the national sample ($N=651$).
  For visibility, only the two largest coefficients in each category are displayed.
  Full results are reported in Table~\ref{tab:lasso_results}.}
  \Description{Diagram divided into two halves, one for GAI (top) and one for human therapists (bottom), with arrows pointing toward intention outcomes.
  Top half (GAI): perceived benefits include Emotional Benefit (beta=0.40) and Personalization (beta=0.16), shown as blue arrows labeled Increase.
  Perceived barriers include Concern of Reliability (beta=-0.12) and Technology Illiteracy (beta=-0.08), shown as red arrows labeled Decrease.
  Therapist panel above divider: benefits Personalization (beta=-0.06) and Emotional Benefit (beta=-0.03) decrease intention (red arrows).
  Barriers Help Seeking Norm (beta=0.09) and Mental Health Illiteracy (beta=0.06) increase intention (blue arrows).
  Bottom half: for GAI, benefits Affordability (beta=-0.06) and Accessibility (beta=-0.03) decrease intention (red arrows), while barriers Technology Illiteracy (beta=0.14) and Mental Health Illiteracy (beta=0.08) increase intention (blue arrows).
  For therapists, benefits Emotional Benefit (beta=0.33) and Affordability (beta=0.24) increase intention (blue arrows).
  Barriers Help Seeking Norm (beta=-0.13) and Concern of Privacy (beta=-0.05) decrease intention (red arrows).
  Right side icons indicate intention to use GAI (robot icon) or human therapist (person icon).
  Key takeaway: in the national sample, emotional and personalization benefits drive GAI uptake, while therapist use is predicted by strong emotional and affordability benefits, offset by help-seeking norms and privacy concerns.}
  \label{fig:lasso_national}
\end{figure}

Figure \ref{fig:lasso_national} illustrates the significant  perceived benefits and barriers for GAI and human therapist in the national sample. 

In the national sample, the most influential predictor for the intention to use GAI was emotional benefit of GAI ($\beta = .40$), marking it as the strongest single factor in the national sample.
This suggests that individuals are drawn to GAI primarily when they perceive it can offer emotional support.
This contrasts with the university sample, where relational benefits were the strongest predictor.
Here, personalization also showed a moderate effect ($\beta = .16$), indicating that the capacity of GAI to adapt to users’ needs is an important motivator across populations.

Technology illiteracy was more salient in the national sample.
It was a significant negative predictor of GAI use ($\beta = -0.08$), suggesting that lack of familiarity or confidence with digital tools may hinder adoption of GAI services more severely among the general population than among university students.
Also, concern about the reliability of GAI was the strongest barrier ($\beta = -0.12$).
Therapists factors also played subtle roles.
Individuals who viewed therapists as providing emotional benefit ($\beta = -0.03$) or personalization ($\beta = -0.06$) were slightly less inclined to choose GAI.

When considering the intention to use therapists, the national sample again highlighted affordability of therapist as a decisive factor ($\beta = .24$).
Emotional benefit was also a strong positive predictor ($\beta = .33$), further reinforcing that emotional responsiveness is a key dimension of perceived therapeutic value.
Notably, technology illiteracy ($\beta = .14$) was associated with higher intention to use therapists, perhaps reflecting a preference for human interaction when digital tools are seen as inaccessible or intimidating.
Additionally, individuals with limited mental health literacy to use GAI, those unsure of what kind of support exists or how GAI could help, were more inclined to rely on therapists instead ($\beta = .08$).
This suggests that unfamiliarity with AI-driven tools may redirect users to more conventional options.

In summary, while both GAI and therapists are seen as viable options for mental health support, their selection hinges on different belief structures.
For GAI, emotional benefits and personalization are key drivers, but concerns about reliability and technological competence remain significant barriers.
For therapists, affordability and emotional benefit remain dominant motivators, particularly in populations less comfortable with or informed about digital mental health tools. 

\section{Discussion}

This study investigated how individuals choose between generative AI (GAI) tools and human therapists for mental health support, using the Health Belief Model (HBM) as a guiding framework.
Drawing on data from both university students and a nationally representative adult sample, and applying LASSO regression to identify salient belief structures, we aimed to uncover the psychological reasoning behind help-seeking preferences.
To our knowledge, this is the first study to empirically compare user perceptions of GAI and human therapists side by side, offering insights that inform the design of interactive systems for mental health support.

Overall, we identified key perceived benefits and barriers associated with each modality.
Interestingly, some beliefs that were highly endorsed did not necessarily play central roles in shaping users’ intentions, suggesting a complex relationship between endorsement and behavioral influence.
Based on these findings, we discuss how insights from this study can inform the design and development of GAI tools for mental health support.
For example, what makes a GAI tool appealing to users?
What potential pitfalls should developers consider when designing AI-driven support systems?
We also outline directions for future research to further investigate how users decide between GAI and human therapists, or whether they choose to engage with both, as part of their mental health journey.

\subsection{Design beyond Accessibility}

Across both the university and national samples, participants highlighted accessibility and affordability as clear advantages of generative AI (GAI).
These perceptions are consistent with prior research documenting how digital tools can circumvent long-standing barriers to mental health care, such as provider shortages, scheduling constraints, and high costs ~\cite{hua2025scoping, sezgin2024behavioral}.
Yet, our analysis revealed that these structural benefits, while acknowledged, did not significantly predict intention to use GAI.
This discrepancy underscores a critical insight: reducing logistical barriers is necessary but not sufficient.
For GAI to become a trusted part of mental health support, it must also demonstrate emotional credibility.

The strongest predictors of GAI intention were not access-related but affective in nature: emotional benefit, relational support, and personalization.
These findings suggest that users look for more than convenience when deciding whether to engage with AI systems; they seek evidence that the system can meet core psychological needs.
Notably, respondents consistently rated GAI lower than therapists in personalization and emotional support.
This is striking because personalization is often promoted as a hallmark of AI systems.
The gap points to a disconnect between technical capacity and subjective user experience: while large language models can generate contextually tailored responses, users remain skeptical about whether these responses translate into genuine care.

This skepticism may stem from how current GAI tools convey empathy.
Simulated warmth or validation may appear formulaic, lacking the adaptive nuance and nonverbal attunement characteristic of human therapists.
Prior work emphasizes that trust in AI-mediated health support grows when systems actively demonstrate responsiveness and sensitivity to user context ~\cite{glickman2025human}.
Our findings align with this perspective, showing that intentions to use GAI rise when users perceive the tool as capable of emotional resonance, rather than merely informational accuracy or logistical efficiency.

The implications for design are clear: future GAI systems must move beyond optimizing for access and instead cultivate emotional presence.
This can be pursued through features such as adaptive dialogue that evolves with user disclosures, mechanisms for validating emotional experiences in real time, and consistency in tone that conveys warmth and attentiveness.
Moreover, integrating feedback loops where users can signal whether responses feel supportive could help bridge the gap between technical outputs and subjective trust.
Such design choices are not cosmetic but central to adoption. Without them, GAI risks being dismissed as efficient yet emotionally hollow.

In short, structural advantages open the door, but emotional benefits sustains engagement.
To meaningfully complement or substitute for traditional therapy, GAI must earn users’ trust by demonstrating that it can “show up” emotionally.
This means treating empathy, personalization, and validation not as optional add-ons but as core design priorities.

\subsection{Match the Tool to the Terrain of Support}
Our findings reveal that perceptions of generative AI (GAI) do not exist in isolation but are shaped by the broader “terrain of support” in which users are embedded.
For university students, who typically have access to counseling centers, peer support networks, and mental health awareness initiatives ~\cite{osborn2022university}, GAI was not positioned as a replacement for therapy.
Instead, it functioned as a complement, a flexible tool to fill gaps between sessions, offer immediacy, or provide low-barrier engagement.
This aligns with prior work showing that digitally literate youth often adopt “assemblage” approaches to care, combining digital and in-person supports to create layered pathways of help-seeking ~\cite{riboldi2024understanding, lupton2021young}.
Within this environment, GAI appears most effective as an adjunct, extending the reach of existing services without displacing them.

In contrast, the national sample, representing a wider range of ages, occupations, and geographic contexts, often approached GAI as a substitute.
Here, intention to use GAI was negatively associated with intention to use therapists.
This pattern suggests that where professional services are less accessible due to financial, geographic, or logistical barriers, individuals may feel compelled to choose between modalities rather than combine them ~\cite{lochner2025digital, mohr2013behavioral}.
For these populations, GAI becomes a primary scaffold for support, valued not only for convenience but as a functional alternative when traditional pathways are unavailable or impractical.
This substitution dynamic is consistent with prior evidence that digital health tools frequently serve as first-line interventions in contexts of scarcity ~\cite{chen2010racial}.

Taken together, these contrasting patterns highlight that adoption is less about the inherent qualities of GAI and more about the ecosystem in which it is situated.
The same tool can flex into different roles depending on users’ social resources and structural constraints.
This insight has important design implications. In resource-rich settings such as universities, GAI should be optimized for complementarity, offering features that extend therapeutic processes, reinforce coping strategies, or prepare users for in-person sessions.
In resource-scarce contexts, however, GAI must be equipped to act as a primary support, requiring greater emphasis on breadth of psychoeducational content, emotional responsiveness, and trust-building safeguards.
Ultimately, the value of GAI lies not in its uniform application but in its adaptability to diverse terrains of support.
Designing systems that can flex between companion and scaffold, depending on whether users are supplementing existing resources or compensating for their absence, will be critical to ensuring GAI’s relevance and equity across populations.

\subsection{Build for Literacy Before Building New Features}

Our findings underscore that adoption of generative AI (GAI) for mental health support is constrained less by what the technology can do and more by whether people feel capable of using it.
In the national sample, barriers such as technological illiteracy and limited mental health literacy strongly shaped preferences, often steering individuals toward traditional therapy despite recognizing GAI’s affordability and accessibility.
This suggests that for many, the decisive factor is not the presence of advanced features but the confidence to navigate the system in the first place.

Prior research on digital health adoption echoes this point: usability and literacy consistently predict engagement more reliably than system sophistication ~\cite{wamala2025digital, mehrotra2024systematic}.
When users struggle to understand how to interact with a system, interpret its outputs, or trust its data practices, even the most innovative features remain underutilized.
Our results reinforce this dynamic. Participants who doubted their digital skills or mental health knowledge reported higher therapist intention, not necessarily because they preferred human care but because they lacked the efficacy to engage with AI-based support.
This highlights an important boundary condition: without scaffolding user literacy, GAI will remain inaccessible to those who might benefit most.

Designing for literacy requires rethinking priorities.
Instead of racing to integrate new capabilities, developers should focus on onboarding flows and educational scaffolds that demystify system use.
For example, transparent explanations of what the tool can and cannot do, interactive tutorials that model effective dialogue, and in-context prompts that guide users through difficult moments can reduce cognitive burden and build user confidence.
Clear communication about data use and privacy protections is equally essential, as uncertainty about information handling amplifies distrust ~\cite{marin2025ethical}.

Importantly, scaffolding literacy is not only about technology but also about mental health understanding.
Users with limited familiarity with psychological terms or coping strategies may need psychoeducational support to interpret AI outputs meaningfully.
Embedding accessible explanations, normalizing mental health concepts, and offering tiered pathways of information can help users make sense of suggestions and apply them effectively.
In this way, literacy-building becomes a foundation for both trust and sustained engagement. In short, before layering on advanced capabilities, GAI systems must lower the cognitive barrier to entry.
Building for literacy first ensures that tools are not only powerful but also usable, equitable, and genuinely supportive for diverse populations.

\subsection{Design for Pathways, Not Silos}

Another key implication is that system design should account for the pathways users construct, rather than assuming linear or siloed choices.
Our findings suggest that individuals do not approach help-seeking through neat “either/or” decision trees.
Instead, they build adaptive and sometimes messy pathways shaped by their resources, trust levels, and literacy.
A user may begin with self-help strategies, shift to GAI when seeking low-barrier support, and eventually engage a therapist for deeper work (or reverse this order depending on circumstances).
Designing as though GAI and therapy are discrete, siloed options risks overlooking the fluid ways people actually navigate care.

Supporting pathways means enabling smooth transitions between self-help, AI-based support, and professional services.
Features such as context-aware referral prompts could encourage users to escalate care when needed, while data-sharing mechanisms (implemented with strict consent and privacy protections) could help therapists build on insights gathered during AI interactions.
Likewise, in-the-moment nudges might help users move between modalities, such as suggesting therapy preparation questions after a series of GAI self-check-ins.
Prior research emphasizes that blended or layered care is often more effective than single-mode interventions ~\cite{cornish2017meeting, thieme2023designing}, and our results reinforce the value of designing systems that can flexibly interconnect rather than compete.
By prioritizing pathways, GAI developers can reframe their tools not as standalone silos but as integral threads within a broader ecosystem of care.
This shift in design perspective acknowledges the reality of users’ lived decision-making and maximizes the likelihood that GAI will be experienced as a bridge rather than a barrier in mental health journeys.

\subsection{Fill the Gaps Therapists Can’t Always Cover}

Our findings reaffirm that therapists remain unmatched in their capacity for depth, trust, and empathy.
Participants consistently rated human therapists higher than GAI in personalization, emotional benefit, and relational support, echoing longstanding evidence that therapeutic alliances are grounded in human attunement and contextual understanding ~\cite{zhang2024can}.
Yet therapy is not without constraints. Structural barriers such as limited appointment availability, high costs, and the natural boundaries of session-based work mean that even the most effective therapists cannot offer continuous, real-time support.

This is where GAI has the potential to complement rather than compete.
By leaning into what human providers cannot always deliver, including immediacy, scalability, and constant availability, GAI can occupy the “between-session” spaces that often go unsupported.
For example, GAI systems could provide daily emotion check-ins, suggest coping strategies in moments of distress, or help clients track progress over time.
These functions extend therapeutic continuity, ensuring that support does not vanish once the session ends.
Prior work on digital mental health tools highlights that continuity of care is critical for maintaining engagement and reinforcing skill practice ~\cite{fitzpatrick-2017, inkster2018empathy}, a role well-suited for AI’s scalable presence.

Importantly, positioning GAI in this complementary space avoids framing it as a rival to therapy.
Rather than replicating the deep relational work of human therapists, GAI can strategically focus on coverage gaps, which include supporting immediacy when crises arise outside of office hours, reinforcing psychoeducation, and offering low-barrier engagement for users hesitant to reach out directly.
When designed with this role in mind, GAI tools can strengthen the overall care ecosystem by catching what might otherwise slip through the cracks.

By filling gaps that therapists cannot always cover, GAI can transform from a perceived substitute into a trusted extension of care.
This requires careful calibration: ensuring emotional benefits and trustworthiness while being transparent about limitations.
Done well, GAI can act as the connective tissue of mental health support, offering continuity without competition and amplifying the therapeutic process rather than diluting it.

\subsection{Limitations and Future Directions}

While this study provides important insights into how individuals consider generative AI (GAI) and human therapists for mental health support, several limitations suggest promising avenues for future research.

First, the cross-sectional design limits causal inference.
Although LASSO regression allowed us to identify belief structures that predict help-seeking intentions, the findings do not indicate how these beliefs develop or shift over time.
Longitudinal or experimental studies are needed to examine how real-world exposure to GAI systems influences user attitudes and behaviors.
For example, tracking individuals before and after engaging with a GAI-based support tool could illuminate whether concerns—such as skepticism about emotional resonance or personalization—diminish with actual use.

Second, the generalizability of our findings is constrained by sample characteristics.
The university sample consisted of students from a single institution with concentrations in psychology and computer science, which may have influenced their attitudes toward both mental health and emerging technologies.
The national sample, although broader, was limited to individuals with internet access and a baseline level of digital literacy.
To improve inclusivity and relevance, future studies should engage more diverse populations, including offline, underserved, and clinical communities who may face distinct barriers or hold different perceptions of mental health technologies.

Third, even though this study is grounded in psychological beliefs—such as the idea that people seek emotional or relational support from GAI tools—we still do not know what specifically promotes the subjective feeling of being “psychologically supported.” Identifying predictors of intention is not the same as uncovering the underlying mechanisms of felt support. Future research should draw on clinically validated concepts such as perceived empathy, therapeutic alliance, expectation of users and patients, and emotion regulation ~\cite{concannon2024measuring, elvins2008conceptualization, greenberg2006patient, berking2008emotion}. For example, it remains unclear whether users interpret AI validation as empathy, whether description of AI tools enhances user's expectation, or whether repeated engagement contributes to emotion regulation. Experimental and qualitative studies could help unpack these processes, offering a more precise understanding of how and why GAI interactions are experienced as supportive.

Finally, while this study identifies beliefs that predict intention to use GAI and therapists, it does not directly reveal the cognitive or behavioral mechanisms underlying decision-making.
Future research should explore how individuals frame their mental health support choices in real-world contexts, particularly under conditions of emotional distress or limited access.
Qualitative or mixed-methods approaches could be especially useful in unpacking how users weigh trade-offs, compare modalities, and define trust or emotional adequacy in technology-mediated care.
Understanding these psychological processes will be essential for designing tools that are not only functional but also meaningful and supportive in users’ lived experiences.

\section{Conclusion}
This study provides important insights into how individuals navigate choices between generative AI (GAI) and human therapists for mental health support.
By integrating the Health Belief Model (HBM) and multi-modality frameworks, we uncovered distinct psychological mechanisms influencing help-seeking behaviors across two populations, university students and a nationally representative adult sample.
Our findings highlight that individuals' perceptions of benefits and barriers significantly shape their intentions toward GAI and therapist modalities, though these perceptions function differently across populations.
University students primarily adopted a complementary approach, leveraging GAI for its emotional responsiveness, relational support, and personalized interactions, while still valuing therapists for deeper emotional and practical engagement.
Conversely, the national sample exhibited a substitution pattern, viewing GAI and therapists as interchangeable resources, with decisions largely driven by affordability, emotional support, and perceived reliability.
These results have meaningful implications for the future design and deployment of GAI tools in mental health contexts.
Developers should prioritize building emotional resonance, trustworthiness, and personalization into AI-driven support systems rather than solely focusing on structural advantages like accessibility or affordability.
Additionally, understanding population-specific dynamics can guide tailored interventions, ensuring both GAI and traditional therapeutic options coexist effectively within mental health ecosystems.
Future research should further explore how demographic factors, digital literacy, and evolving technology perceptions shape these complex decisions, ultimately informing more nuanced integration strategies for GAI in mental health care.

\section{Acknowledgments}

\begin{acks}
This research was supported by the Department of Psychology at the University of Florida through the Trish Calvert Ring Graduate Research Award.
\end{acks}

\bibliographystyle{ACM-Reference-Format}
\bibliography{sample-base}

\end{document}